\documentclass[sigconf,nonacm]{acmart}

\AtBeginDocument{%
  }
\usepackage{amsmath,amsfonts}
\usepackage{algorithmic}
\usepackage[binary-units=true]{siunitx}
\usepackage{graphicx}
\usepackage{textcomp}
\usepackage{bmpsize}
\usepackage{xcolor}
\usepackage{lipsum}
\usepackage{glossaries}
\usepackage{booktabs}
\usepackage[shortlabels]{enumitem}
\usepackage{makecell}
\usepackage{subcaption}
\usepackage[english]{babel}
\usepackage[T1]{fontenc}
\newcommand\blfootnote[1]{%
  \begingroup
  \renewcommand\thefootnote{}\footnote{#1}%
  \addtocounter{footnote}{-1}%
  \endgroup
}

\usepackage{cleveref}

\usepackage{hyperref}

\DeclareSIUnit[per-mode=symbol]\bps{\bit\per\second}
\DeclareSIUnit[per-mode=symbol]\kbps{\kilo\bps}
\DeclareSIUnit[per-mode=symbol]\Mbps{\mega\bps}
\DeclareSIUnit[per-mode=symbol]\Gbps{\giga\bps}
\DeclareSIUnit[per-mode=symbol]\nanosec{\nano\second}
\DeclareSIUnit[per-mode=symbol]\packet{packet}
\DeclareSIUnit[per-mode=symbol]\packetps{\packet\per\second}
\DeclareSIUnit\microsec{\SIUnitSymbolMicro s}
\DeclareSIUnit\byte{B}
\DeclareSIUnit\bit{bit}
\DeclareSIUnit\terabyte{TB}

\setcopyright{acmlicensed}
\copyrightyear{2018}
\acmYear{2018}
\acmDOI{XXXXXXX.XXXXXXX}
\begin{document}

%%
%% The "title" command has an optional parameter,
%% allowing the author to define a "short title" to be used in page headers.
\title{Narrowing the Gap between TEEs Threat Model and Deployment Strategies}

%%
%% The "author" command and its associated commands are used to define
%% the authors and their affiliations.
%% Of note is the shared affiliation of the first two authors, and the
%% "authornote" and "authornotemark" commands
%% used to denote shared contribution to the research.
\author{Filip Rezabek}
\affiliation{%
  \institution{Flashbots\\Technical University of Munich}
  \country{Germany}
}

\author{Jonathan Passerat-Palmbach}
\affiliation{%
  \institution{Flashbots\\Imperial College London}
   \country{UK}
}
\author{Moe Mahhouk}
\affiliation{%
  \institution{Flashbots}
  \country{Germany}
}

\author{Frieder Erdmann}
\affiliation{%
  \institution{Flashbots}
  \country{Germany}
}

\author{Andrew Miller}
\affiliation{%
  \institution{Flashbots}
  \country{USA}
}
%%
%% By default, the full list of authors will be used in the page
%% headers. Often, this list is too long, and will overlap
%% other information printed in the page headers. This command allows
%% the author to define a more concise list
%% of authors' names for this purpose.
\renewcommand{\shortauthors}{Rezabek et al.}

%%
%% The abstract is a short summary of the work to be presented in the
%% article.
\begin{abstract}
Confidential Virtual Machines (CVMs) provide isolation guarantees for data in use, but their threat model does not include physical level protection and side-channel attacks. Therefore, current deployments rely on trusted cloud providers to host the CVMs' underlying infrastructure. However, TEE attestations do not provide information about the operator hosting a CVM. Without knowing whether a Trusted Execution Environment (TEE) runs within a provider's infrastructure, a user cannot accurately assess the risks of physical attacks. We observe a misalignment in the threat model where the workloads are protected against other tenants but do not offer end-to-end security assurances to external users without relying on cloud providers. The attestation should be extended to bind the CVM with the provider. A possible solution can rely on the Protected Platform Identifier (PPID), a unique CPU identifier. However, the implementation details of various TEE manufacturers, attestation flows, and providers vary. This makes verification of attestations, ease of migration, and building applications without relying on a trusted party challenging, highlighting a key limitation that must be addressed for the adoption of CVMs. We discuss two points focusing on hardening and extensions of TEEs' attestation.
\end{abstract}

%%
%% The code below is generated by the tool at http://dl.acm.org/ccs.cfm.
%% Please copy and paste the code instead of the example below.
%%

%%
%% Keywords. The author(s) should pick words that accurately describe
%% the work being presented. Separate the keywords with commas.
%% A "teaser" image appears between the author and affiliation
%% information and the body of the document, and typically spans the
%% page.

\received[Accepted]{to the 8th Edition of the System Software for Trusted Execution (SysTEX) '25 Workshop, co-located with EuroS\&P '25.}

%%
%% This command processes the author and affiliation and title
%% information and builds the first part of the formatted document.
\maketitle

\blfootnote{This paper was originally accepted to the 8th Edition of the System Software for Trusted Execution (SysTEX) '25 Workshop, co-located with EuroS\&P '25.}
\newacronym{nic}{NIC}{Network Interface Card}
\newacronym{iiot}{IIoT}{Industrial Internet of Things}
\newacronym{iot}{IoT}{Internet of Things}
\newacronym{cots}{COTS}{Commercial off-the-Shelf}
\newacronym{rtt}{RTT}{Round Trip Time}
\newacronym{e2e}{E2E}{End-to-End}
\newacronym{p2p}{P2P}{Peer-to-Peer}
\newacronym{gptp}{gPTP}{generic Precision Time Protocol}
\newacronym{phc}{PHC}{PTP Hardware Clock}
\newacronym{gm}{GM}{Grandmaster Clock}
\newacronym{tc}{tc}{traffic control}
\newacronym[plural=TCLs,firstplural=traffic classes (TCLs)]{tcl}{TCL}{Traffic Class}
\newacronym{qos}{QoS}{Quality of Service}
\newacronym{ecdf}{ECDF}{Empirical Cumulative Distribution Function}
\newacronym{be}{BE}{Best Effort}
\newacronym{kpi}{KPI}{Key Performance Indicator}
\newacronym{skb}{SKB}{Socket Buffer}
\newacronym{sut}{SUT}{System Under Test}
\newacronym{phy}{PHY}{Physical Layer}
\newacronym{udp}{UDP}{User Datagram Protocol}
\newacronym{bmca}{BMCA}{Best Master Clock Algorithm}
\newacronym{tcp}{TCP}{Transmission Control Protocol}
\newacronym{os}{OS}{Operating System}
\newacronym{irq}{IRQ}{Interrupt Request}
\newacronym{cpu}{CPU}{Central Processing Unit}
\newacronym{smp}{SMP}{Symmetrical Multiprocessing}
\newacronym{smt}{SMT}{Simultaneous Multi-Threading}
\newacronym{rt}{RT}{Real-Time}
\newacronym{hw}{HW}{hardware}
\newacronym{sw}{SW}{software}
\newacronym{kc}{KC}{Key Contribution}
\newacronym{pcap}{PCAP}{Packet Capture}
\newacronym{utc}{UTC}{Coordinated Universal Time}
\newacronym{tai}{TAI}{International Atomic Time}
\newacronym{txtime}{TxTime}{transmission time}
\newacronym{macsec}{MACsec}{Media Access Control Security}
\newacronym{hpc}{HPC}{High Performance Computer}
\newacronym{lpc}{LPC}{Low Performance Computer}
\newacronym{engine}{EnGINE}{Environment for Generic In-vehicular Networking Experiments}
\newacronym{sc}{SC}{Secure Channel}
\newacronym{fpga}{FPGA}{Field Programmable Gate Array}
\newacronym{fqcodel}{FQ\_CoDel}{Fair Queuing with Controlled Delay}
\newacronym{mac}{MAC}{Media Access Control}
\newacronym{ip}{IP}{Internet Protocol}
\newacronym{ws}{WS}{window size}
\newacronym{ram}{RAM}{Random-Access Memory}
\newacronym{is}{IS}{Interframe Spacing}
\newacronym{god}{GOD}{Guaranteed Output Delivery}
\newacronym{pp}{PP}{Payment Processor}
\newacronym{dlp}{DLP}{Discrete Logarithm Problem}
\newacronym{saas}{SaaS}{Software-as-a-Service}
\newacronym{dsa}{DSA}{Digital Signature Algorithm}
\newacronym{ecdsa}{ECDSA}{Elliptic Curve Digital Signature Algorithm}
\newacronym{eddsa}{EdDSA}{Edwards-curve DSA}
\newacronym{dkg}{DKG}{Distributed Key Generation}
\newacronym{zkp}{ZKP}{Zero-Knowledge Proof}
\newacronym{PoC}{PoC}{proof-of-concept}
\newacronym{mpc}{MPC}{Multiparty Computation}
\newacronym{ot}{OT}{Oblivious Transfer}
\newacronym{uc}{UC}{Universal Composability}
\newacronym{vCPUs}{vCPUs}{virtual CPUs}
\newacronym{tps}{TPS}{Transactions Per Second}
\newacronym{tee}{TEE}{Trusted Execution Environment}
\newacronym{tpm}{TPM}{Trusted Platform Module}
\newacronym{vm}{VM}{Virtual Machine}
\newacronym{tdx}{TDX}{Trust Domain Extensions}
\newacronym{sev}{SEV}{Secure Encrypted Virtualization}
\newacronym{snp}{SNP}{Secure Nested Paging}
\newacronym{sgx}{SGX}{Software Guard Extensions}
\newacronym{qemu}{QEMU}{Quick Emulator}
\newacronym{kvm}{KVM}{Kernel-based Virtual Machine}
\newacronym{tsn}{TSN}{Time Sensitive Networking}
\newacronym{methoda}{METHODA}{Multilayer Environment and Toolchain for Holistic NetwOrk Design and Analysis}
\newacronym{pos}{pos}{plain orchestrating service}
\newacronym{tcb}{TCB}{Trusted Computing Base}
\newacronym{dma}{DMA}{Direct Memory Access}
\newacronym{td}{TD}{Trust Domain}
\newacronym{seam}{SEAM}{Secure Arbitration Mode}
\newacronym{maccode}{MAC}{Message Authentication Code}
\newacronym{gcp}{GCP}{Google Cloud Platform}
\newacronym{sme}{SME}{AMD Secure Memory Encryption}
\newacronym{asp}{ASP}{AMD Secure Processor}
\newacronym{psp}{PSP}{AMD Platform Security Processor}
\newacronym{es}{ES}{Encrypted State}
\newacronym{sota}{SotA}{State of the Art}
\newacronym{poc}{PoC}{Proof of Concept}
\newacronym{vlek}{VLEK}{Verified Launch Enclave Key}
\newacronym{vcek}{VCEK}{Verified Chip Endorsement Key}
\newacronym{vmrk}{VMRK}{VM Root Key}
\newacronym{kds}{KDS}{Key Distribution Server}
\newacronym{ark}{ARK}{AMD Root Key}
\newacronym{ask}{ASK}{AMD SEV Key}
\newacronym{qgs}{QGS}{quote generation service}
\newacronym{pccs}{PCCS}{Provisioning Certification Caching Service}
\newacronym{pcs}{PCS}{Intel Provisioning Certification Service}
\newacronym{mpa}{MPA}{Multi-package Registration Agent}
\newacronym{mktme}{MKTME}{Multi-key Total Memory Encryption}
\newacronym{vmm}{VMM}{VM Manager}
\newacronym{tls}{TLS}{Transport Layer Security}
\newacronym{pek}{PEK}{Platform Endorsement Key}
\newacronym{csr}{CSR}{Certificate Signing Request}
\newacronym{pckcert}{PCKC}{Provisioning Certification Key Certificate}
\newacronym{pce}{PCE}{Provisioning Certificate Enclave}
\newacronym{pck}{PCK}{Provisioning Certification Key}
\newacronym{qsk}{QSK}{Quote Signing Key}
\newacronym{tdqe}{TDQE}{TD Quoting Enclave}
\newacronym{ak}{AK}{Attestation Key}
\newacronym{svn}{SVN}{Security Version Number}
\newacronym{r3aal}{R3AAL}{Ring3 Attestation Abstraction Library}
\newacronym{tdqd}{TDQD}{TD Quote Driver}
\newacronym{qgl}{QGL}{Quote Generation Library}
\newacronym{qvl}{QVL}{Quote Verification Library}
\newacronym{qve}{QVE}{Quote Verification Enclave}
\newacronym{crl}{CRL}{Certificate Revocation List}
\newacronym{spd}{SPD}{Seriel Presence Detect}
\newacronym{rmp}{RMP}{Reverse Map Table}
\newacronym{xts}{XTS}{XEX-based Tweaked CodeBook Mode with Ciphertext Stealing}
\newacronym{xex}{XEX}{XOR-Encrypt-XOR}
\newacronym{aes}{AES}{Advanced Encryption Standard}
\newacronym{zk}{ZK}{Zero-knowledge}
\newacronym{ecc}{ECC}{Elliptic Curve Cryptography}
\newacronym{rsa}{RSA}{Rivest–Shamir–Adleman}
\newacronym{nist}{NIST}{National Institute of Standards and Technology}
\newacronym{bls}{BLS}{Boneh-Lynn-Shacham}
\newacronym{smi}{SMI}{System Management Interrupt}
\newacronym{smm}{SMM}{System Management Mode}
\newacronym{vmpl}{VMPL}{Virtual Machine Protection Level}
\newacronym{mmio}{MMIO}{Memory-mapped I/O}
\newacronym{svsm}{SVSM}{Secure VM Service Module}
\newacronym{vtpm}{vTPM}{virtual Trusted Platform Module}
\newacronym{vmx}{VMX}{Virtual Machines Extension}
\newacronym{cvm}{CVM}{Confidential VM}
\newacronym{tcg}{TCG}{Trusted Computing Group}
\newacronym{rtmr}{RTMR}{Runtime Extendable Measurement Register}
\newacronym{mrtd}{MRTD}{Measurement of Trust Domain}
\newacronym{ppid}{PPID}{Protected Platform Identifier}
\newacronym{uuid}{UUID}{Universally Unique Identifier}
\newacronym{pki}{PKI}{Public Key Infrastructure}
\newacronym{as}{AS}{Autonomous System}
\newacronym{cca}{CCA}{Confidential Compute Architecture}
\newacronym{vtl}{VTL}{Virtual Trust Level}
\newacronym{dcap}{DCAP}{Data Center Attestation Primitives}

\newcommand{\encircled}[2][0.8mm]{%
    \raisebox{.5pt}{%
        \textcircled{%
            \raisebox{0.35pt}{%
                \kern #1
                \scalebox{0.70}{#2}
            }%
        }%
    }%
}
\makeatletter
\newcommand*{\ensquared}[1]{\relax\ifmmode\mathpalette\@ensquared@math{#1}\else\@ensquared{#1}\fi}
\newcommand*{\@ensquared@math}[2]{\@ensquared{$\m@th#1#2$}}
\newcommand*{\@ensquared}[1]{%
\tikz[baseline,anchor=base]{\node[draw,outer sep=0pt,inner sep=0.6mm,minimum width=4mm] {#1};}} 
\makeatother

\definecolor{ourgreen}{rgb}{0.00,0.49,0.19}
\definecolor{ourred}{rgb}{0.77,0.03,0.11}
\definecolor{ourorange}{rgb}{0.89,0.45,0.13}
\definecolor{ourgrey}{rgb}{0.60,0.60,0.60}
\def\yes{\textcolor{ourgreen}{\large\checkmark}}
\def\maybe{\textcolor{ourorange}{\Large$\circ$}} % $\mathbit
\def\no{\textcolor{ourred}{\Large\texttimes}}
\def\unknown{\textcolor{ourgrey}{\encircled[1mm]{?}}}
\glsdisablehyper

\section{Introduction}
\label{sec:intro}
Many applications require safeguarding sensitive data during processing. 
While traditional security measures protect data at rest and in transit, data in use remains vulnerable to threats like unauthorized access and malicious insiders~\cite{gramineTDX-dmitrii2024,ménétrey2022exploratory}.
This risk is especially concerning in cloud environments, where multiple tenants share physical hardware on (ideally) untrusted third-party infrastructure, heightening data breach chances. 
Confidential Computing addresses these challenges by protecting data in use through \glspl{tee}.
\glspl{tee} provide an isolated environment where sensitive computations can be executed without interference, even from higher-privileged software like operating systems or hypervisors. 
Recently, technologies such as Intel \gls{tdx} or AMD \gls{sev}-\gls{snp} run the whole \gls{vm} in such an isolated environment. 
As part of their \gls{tcb}, they include the guest OS and guest \gls{vm}'s privileged users, but can protect against malicious host \gls{os} or hypervisor~\cite{gramineTDX-dmitrii2024}.
External users can request a hardware-signed attestation report to verify that key components remain untampered, including code and data inside the \gls{tee}. 
This combination of isolation and attestation enables secure execution of sensitive workloads, even in untrusted environments.
However, both AMD~\cite{SideChannels-Mengyun2022} and Intel~\cite{gramineTDX-dmitrii2024} exclude memory integrity~\cite{247672}, side channels and sophisticated physical attacks e.g., microscope probing or fault injection, from their threat models. 
While most side channels can be mitigated at the application level, physical attack vectors require users to trust the physical host location of the \glspl{cvm}.
This is especially needed for use cases in the Web3 space, where \glspl{tee} could protect millions of dollars in value~\cite{8806762,Rabimba_2021}.
The robustness of \glspl{cvm} depends on ensuring they operate in a trusted environment, relying on operators who do not tamper with nodes and enforce strong access policies.
Remote attestation verifies that communication terminates in a \gls{tee} on an authenticated platform but does not provide details about the operational environment.

Therefore, the attestation flow should be extended with additional information, assuring that the environment where the \gls{tee} platform runs is a trusted cloud data center, thus strengthening the relation to the provider. 
This closes the gap between the threat model of current \glspl{tee} and the trust in the infrastructure owner. 
A recently proposed solution called LooseSEAL relies on the \gls{ppid} to derive keys originating from \glspl{cvm} on the same machine~\cite{LooseSEAL-2024}. 
The \gls{ppid} is generated based on the \gls{uuid} \cite{Intel-specs-API} of Intel or the CPU\_ID \cite{AMD-specs-abi} of AMD CPUs that run inside the cloud's infrastructure.
As this feature is currently not implemented, the provider must enhance the attestation capabilities. 
Besides, each provider and ideally \gls{tee} manufacturers should implement the same flow for ease of migration, requiring industry standardization. 
Another option is to rely on a certification party that certifies the physical location to ensure the usage of untampered hardware and additional intrusion detection to detect physical access to the devices, as is the case of Apple~\cite{apple_pcc_hardware_integrity}. 
Another, even more demanding, approach is to extend the threat model to include physical attacks and tampering with the chip manufacturer's supply chain.
However, this is less likely as it requires new \gls{tee} designs~\cite{flashbots-blog-TTEE}.

Our work aims to bring discussion points (DPs) about:
\begin{enumerate}[label={\bfseries DP\arabic*}, leftmargin=0.87cm]
    \item Unification/standardization of \gls{ppid} \& deployments.
    \item Threat model extension by physical access.
\end{enumerate}

\section{TEEs and Attestation Flows}
\begin{figure}
    \centering
    \begin{subfigure}{0.49\columnwidth}
    \centering
    \includegraphics[width=.5\columnwidth]{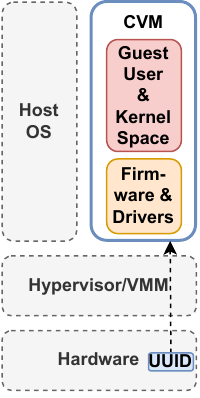}
    \caption{Bare/CVM Flow}
    \label{fig:bare-flow}
    \end{subfigure}
    \begin{subfigure}{.49\columnwidth}
    \centering
    \includegraphics[width=.73\columnwidth]{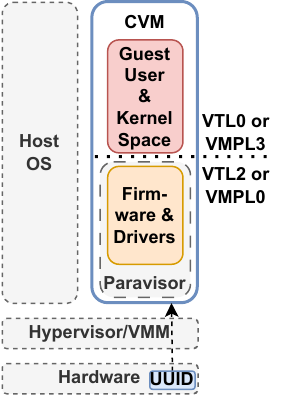}
    \caption{Paravirtualization Flow}
    \label{fig:para-flow}
    \end{subfigure}
    \caption{Simplified TEE Attestation Flow for various Deployments. The provider controls the grey dotted boxes.}
    \label{fig:attest-flows}
\end{figure}

We introduce relevant background information supporting the \textbf{DPs}. 
\gls{vm}-based \glspl{tee}, such as Intel \gls{tdx}~\cite{inteltdx3:online,9448036} or AMD \gls{sev}-\gls{snp}~\cite{ménétrey2022exploratory,AMDSEVGit,li2022sokteeassistedconfidentialsmart}, enhance \gls{vm} security through encrypting and isolating guest \glspl{vm} from the hypervisor and supporting nested virtualization. 
Users gain confidence in a given TEE enclave via the request of a remote attestation. 
For \glspl{cvm},~\Cref{fig:attest-flows} presents two attestation flows varying between bare metal/native virtualization (\ref{fig:bare-flow}) and with an additional paravirtualized layer (\ref{fig:para-flow}) and how is UUID available to \gls{cvm}. 
The paravisor allows for live migration of the \gls{cvm} and provides an additional layer of virtual drivers between the guest OS and underlying \gls{vmm}.
In the bare metal setup (\Cref{fig:bare-flow}), the \gls{cvm} runs directly on the hypervisor, e.g., QEMU. 
Attestation in this scenario involves verifying the firmware, operating system, and \gls{tee} itself. 
On the other hand, in the paravirtualized environment (\Cref{fig:para-flow}), the \gls{cvm} additionally relies on a paravisor, e.g., OpenHCL~\cite{microsoft2025Feb} or COCONUT~\cite{coconut-svsm2025Feb}.
The paravisor implements an access mode present as a \gls{vtl} for Intel \gls{tdx} and \gls{vmpl} for AMD \gls{sev}-\gls{snp}. 
Of note, \gls{vmpl}0 is the highest privilege level, and \gls{vtl}0 is the lowest, hinting at other implementation approaches.
The attestation report should include verification of the same components as in a regular deployment and the paravirtualization stack. 
This requires the paravisor's components to be open-source to enable reproducible builds and thus obtain the checksum to compare with the value in the attestation's fields. 
This is, however, not always the case, as was the case of Microsoft Azure's paravisor before OpenHCL~\cite{microsoftConfidentialAzure}. 
Even when using the paravisor approach, the quote contains the \gls{ppid} constructed during Intel's initial platform verification. 
For the case of live migration, the verifier must regularly be made aware of migration or request attestation, as the \gls{ppid} is hardware-dependent.

\section{TEE Attestation Extensions \& Beyond}
Building on top of the background information, we address the \textbf{DPs} introduced in \Cref{sec:intro}.
The \textbf{DP1} focuses on the solution's reliance on \gls{ppid} and \textbf{DP2} motivates physical attacks integration to the \gls{tee} threat model.
One way cloud providers can offer assurance that a given \gls{tee} is indeed cloud-based could be to leverage \gls{ppid}.
The \gls{ppid} is a unique identifier derived uniquely for each Intel CPUs, allowing attestation reports to be linked to a known and verifiable machine or infrastructure for Intel TDX and SGX.
In the case of \gls{dcap} attestation quotes, the Platform ID (first \SI{16}{\byte} of user data) is either the encrypted \gls{ppid} or is derived from it. 
The \gls{ppid} is encrypted using Intel's public key of the \gls{pcs}, with the private key being only owned by Intel. 
Therefore, to modify the \gls{ppid}, the CPU manufacturer would have to be involved, serving as a separation of interests.
Overall, the \gls{ppid} is a consistent identifier that links attestation quotes to specific physical CPUs.
For the solution to work, the cloud provider must keep a list of its publicly available hardware identifiers so the users can then validate the provided \gls{ppid}.
Such a mechanism would ensure that workloads are executed on certified hardware and within a secure infrastructure. 
We rely on cloud providers for physical protection, so enabling the \gls{ppid} does not increase the attack surface, and as an operator of \gls{cvm}, we can verify the information is correct.
Of note, different \glspl{cvm} on the same hardware share the same \gls{ppid}.
To ensure privacy, we can use a \gls{zk} proof of the attestation and the \gls{ppid} details proving that our node runs in a particular cloud, without disclosing which. 
There are already \gls{zk} instantiations for \gls{dcap}~\cite{automata_dcap_attestation} which can be extended for this setting. 
The effectiveness of this solution depends on how cloud providers implement and disclose such identifiers.
Implementing \glspl{ppid} should also be unified to allow easy migration across providers to mitigate possible friction.
\gls{ppid} is a robust solution considering the current setting and not too demanding from the cloud provider's perspective. 
Nevertheless, the trust in the cloud provider does not increase, as we rely on the provider for physical attack protection. 

However, several challenges must be addressed in designing and deploying such a solution.
One key difficulty is accounting for hardware diversity, as different processor architectures (e.g., Intel \gls{tdx}, AMD \gls{sev}, ARM \gls{cca}) implement \glspl{tee} with varying security models and attestation mechanisms. 
A standardized solution must accommodate these differences while maintaining security guarantees. 
The key issue of physical and side-channel attacks persists. 
Therefore, extending the threat model to include more physical, supply-chain, and side-channels attacks will improve the potential of \gls{tee} as a technology. 

\section{Next Steps for TEEs}
We highlight the need to extend \glspl{tee}' threat model to include physical access attacks. 
Current VM-based \glspl{tee} implicitly trust the cloud provider, which is misaligned with attestation flows that do not bind the provider to the report. 
Available solutions such as \gls{ppid} can improve \glspl{tee} adoption.
Strengthening the threat model and closing the attestation gap requires collaboration among manufacturers, service providers, and researchers. 

%\newpage
% \renewcommand*{\bibfont}{\footnotesize}
% \printbibliography
\bibliographystyle{ACM-Reference-Format}
\bibliography{sample-base}

%%% -*-BibTeX-*-
%%% Do NOT edit. File created by BibTeX with style
%%% ACM-Reference-Format-Journals [18-Jan-2012].

\begin{thebibliography}{19}

%%% ====================================================================
%%% NOTE TO THE USER: you can override these defaults by providing
%%% customized versions of any of these macros before the \bibliography
%%% command.  Each of them MUST provide its own final punctuation,
%%% except for \shownote{} and \showURL{}.  The latter two
%%% do not use final punctuation, in order to avoid confusing it with
%%% the Web address.
%%%
%%% To suppress output of a particular field, define its macro to expand
%%% to an empty string, or better, \unskip, like this:
%%%
%%% \newcommand{\showURL}[1]{\unskip}   % LaTeX syntax
%%%
%%% \def \showURL #1{\unskip}           % plain TeX syntax
%%%
%%% ====================================================================

\ifx \showCODEN    \undefined \def \showCODEN     #1{\unskip}     \fi
\ifx \showISBNx    \undefined \def \showISBNx     #1{\unskip}     \fi
\ifx \showISBNxiii \undefined \def \showISBNxiii  #1{\unskip}     \fi
\ifx \showISSN     \undefined \def \showISSN      #1{\unskip}     \fi
\ifx \showLCCN     \undefined \def \showLCCN      #1{\unskip}     \fi
\ifx \shownote     \undefined \def \shownote      #1{#1}          \fi
\ifx \showarticletitle \undefined \def \showarticletitle #1{#1}   \fi
\ifx \showURL      \undefined \def \showURL       {\relax}        \fi
% The following commands are used for tagged output and should be
% invisible to TeX
\providecommand\bibfield[2]{#2}
\providecommand\bibinfo[2]{#2}
\providecommand\natexlab[1]{#1}
\providecommand\showeprint[2][]{arXiv:#2}

\bibitem[AMD(2022)]%
        {AMDSEVGit}
\bibfield{author}{\bibinfo{person}{AMD}.} \bibinfo{year}{2022}\natexlab{}.
\newblock \bibinfo{booktitle}{\emph{GitHub - AMDESE/AMDSEV: AMD Secure Encrypted Virtualization}}.
\newblock
\newblock
\shownote{(Accessed on 10/15/2023)}.


\bibitem[AMD(2025)]%
        {AMD-specs-abi}
\bibfield{author}{\bibinfo{person}{AMD}.} \bibinfo{year}{2025}\natexlab{}.
\newblock \bibinfo{title}{SEV Secure Nested Paging Firmware ABI Specification}.
\newblock
\urldef\tempurl%
\url{https://www.amd.com/content/dam/amd/en/documents/epyc-technical-docs/specifications/56860.pdf}
\showURL{%
\tempurl}
\newblock
\shownote{[Online; accessed 14. Feb. 2025]}.


\bibitem[{Apple Inc.}({[n.\,d.]})]%
        {apple_pcc_hardware_integrity}
\bibfield{author}{\bibinfo{person}{{Apple Inc.}}} \bibinfo{year}{[n.\,d.]}\natexlab{}.
\newblock \bibinfo{title}{Hardware Integrity in Private Cloud Compute}.
\newblock \bibinfo{howpublished}{\url{https://security.apple.com/documentation/private-cloud-compute/hardwareintegrity}}.
\newblock
\newblock
\shownote{Accessed: 2025-04-01}.


\bibitem[Cheng et~al\mbox{.}(2019)]%
        {8806762}
\bibfield{author}{\bibinfo{person}{Raymond Cheng}, \bibinfo{person}{Fan Zhang}, \bibinfo{person}{Jernej Kos}, \bibinfo{person}{Warren He}, \bibinfo{person}{Nicholas Hynes}, \bibinfo{person}{Noah Johnson}, \bibinfo{person}{Ari Juels}, \bibinfo{person}{Andrew Miller}, {and} \bibinfo{person}{Dawn Song}.} \bibinfo{year}{2019}\natexlab{}.
\newblock \showarticletitle{Ekiden: A Platform for Confidentiality-Preserving, Trustworthy, and Performant Smart Contracts}. In \bibinfo{booktitle}{\emph{2019 IEEE European Symposium on Security and Privacy (EuroS\&P)}}. \bibinfo{pages}{185--200}.
\newblock
\href{https://doi.org/10.1109/EuroSP.2019.00023}{doi:\nolinkurl{10.1109/EuroSP.2019.00023}}


\bibitem[Intel(2024a)]%
        {Intel-specs-API}
\bibfield{author}{\bibinfo{person}{Intel}.} \bibinfo{year}{2024}\natexlab{a}.
\newblock \bibinfo{title}{Intel TDX DCAP: Quote Generation Library and Quote Verification Library}.
\newblock
\urldef\tempurl%
\url{https://download.01.org/intel-sgx/latest/dcap-latest/linux/docs/Intel_TDX_DCAP_Quoting_Library_API.pdf}
\showURL{%
\tempurl}
\newblock
\shownote{[Online; accessed 14. Feb. 2025]}.


\bibitem[Intel(2024b)]%
        {inteltdx3:online}
\bibfield{author}{\bibinfo{person}{Intel}.} \bibinfo{year}{2024}\natexlab{b}.
\newblock \bibinfo{booktitle}{\emph{intel/tdx-module}}.
\newblock
\newblock
\shownote{(Accessed on 05/10/2024)}.


\bibitem[Kilbourn(2024)]%
        {flashbots-blog-TTEE}
\bibfield{author}{\bibinfo{person}{Quintus Kilbourn}.} \bibinfo{year}{2024}\natexlab{}.
\newblock \bibinfo{title}{{Zero Trust Execution Environments - TEE - Trusted Execution Environment / Trustless TEEs - The Flashbots Collective}}.
\newblock
\urldef\tempurl%
\url{https://collective.flashbots.net/t/zero-trust-execution-environments/3966}
\showURL{%
\tempurl}
\newblock
\shownote{[Online; accessed 15. Feb. 2025]}.


\bibitem[Kuvaiskii et~al\mbox{.}(2024)]%
        {gramineTDX-dmitrii2024}
\bibfield{author}{\bibinfo{person}{Dmitrii Kuvaiskii}, \bibinfo{person}{Dimitrios Stavrakakis}, \bibinfo{person}{Kailun Qin}, \bibinfo{person}{Cedric Xing}, \bibinfo{person}{Pramod Bhatotia}, {and} \bibinfo{person}{Mona Vij}.} \bibinfo{year}{2024}\natexlab{}.
\newblock \showarticletitle{Gramine-TDX: A Lightweight OS Kernel for Confidential VMs}. In \bibinfo{booktitle}{\emph{Proceedings of the 2024 on ACM SIGSAC Conference on Computer and Communications Security}} (Salt Lake City, UT, USA) \emph{(\bibinfo{series}{CCS '24})}. \bibinfo{publisher}{Association for Computing Machinery}, \bibinfo{address}{New York, NY, USA}, \bibinfo{pages}{4598–4612}.
\newblock
\showISBNx{9798400706363}
\href{https://doi.org/10.1145/3658644.3690323}{doi:\nolinkurl{10.1145/3658644.3690323}}


\bibitem[Lee et~al\mbox{.}(2020)]%
        {247672}
\bibfield{author}{\bibinfo{person}{Dayeol Lee}, \bibinfo{person}{Dongha Jung}, \bibinfo{person}{Ian~T. Fang}, \bibinfo{person}{Chia che Tsai}, {and} \bibinfo{person}{Raluca~Ada Popa}.} \bibinfo{year}{2020}\natexlab{}.
\newblock \showarticletitle{An {Off-Chip} Attack on Hardware Enclaves via the Memory Bus}. In \bibinfo{booktitle}{\emph{29th USENIX Security Symposium (USENIX Security 20)}}. \bibinfo{publisher}{USENIX Association}, \bibinfo{pages}{487--504}.
\newblock
\showISBNx{978-1-939133-17-5}
\urldef\tempurl%
\url{https://www.usenix.org/conference/usenixsecurity20/presentation/lee-dayeol}
\showURL{%
\tempurl}


\bibitem[Li et~al\mbox{.}(2022b)]%
        {SideChannels-Mengyun2022}
\bibfield{author}{\bibinfo{person}{Mengyuan Li}, \bibinfo{person}{Luca Wilke}, \bibinfo{person}{Jan Wichelmann}, \bibinfo{person}{Thomas Eisenbarth}, \bibinfo{person}{Radu Teodorescu}, {and} \bibinfo{person}{Yinqian Zhang}.} \bibinfo{year}{2022}\natexlab{b}.
\newblock \showarticletitle{A Systematic Look at Ciphertext Side Channels on AMD SEV-SNP}. In \bibinfo{booktitle}{\emph{2022 IEEE Symposium on Security and Privacy (SP)}}. \bibinfo{pages}{337--351}.
\newblock
\href{https://doi.org/10.1109/SP46214.2022.9833768}{doi:\nolinkurl{10.1109/SP46214.2022.9833768}}


\bibitem[Li et~al\mbox{.}(2022a)]%
        {li2022sokteeassistedconfidentialsmart}
\bibfield{author}{\bibinfo{person}{Rujia Li}, \bibinfo{person}{Qin Wang}, \bibinfo{person}{Qi Wang}, \bibinfo{person}{David Galindo}, {and} \bibinfo{person}{Mark Ryan}.} \bibinfo{year}{2022}\natexlab{a}.
\newblock \bibinfo{title}{SoK: TEE-assisted Confidential Smart Contract}.
\newblock
\showeprint[arxiv]{2203.08548}~[cs.CR]
\urldef\tempurl%
\url{https://arxiv.org/abs/2203.08548}
\showURL{%
\tempurl}


\bibitem[Mahhouk(2024)]%
        {LooseSEAL-2024}
\bibfield{author}{\bibinfo{person}{Moe Mahhouk}.} \bibinfo{year}{2024}\natexlab{}.
\newblock \bibinfo{title}{{Loose SEAL: Enabling Crash-Tolerant TDX Applications by Utilizing SGX Sealing Provider Sidecar - TEE - Trusted Execution Environment - The Flashbots Collective}}.
\newblock
\urldef\tempurl%
\url{https://collective.flashbots.net/t/loose-seal-enabling-crash-tolerant-tdx-applications-by-utilizing-sgx-sealing-provider-sidecar/4243/1}
\showURL{%
\tempurl}
\newblock
\shownote{[Online; accessed 14. Feb. 2025]}.


\bibitem[Microsoft(2025)]%
        {microsoft2025Feb}
\bibfield{author}{\bibinfo{person}{Microsoft}.} \bibinfo{year}{2025}\natexlab{}.
\newblock \bibinfo{title}{{openvmm}}.
\newblock
\urldef\tempurl%
\url{https://github.com/microsoft/openvmm}
\showURL{%
\tempurl}
\newblock
\shownote{[Online; accessed 14. Feb. 2025]}.


\bibitem[Ménétrey et~al\mbox{.}(2022)]%
        {ménétrey2022exploratory}
\bibfield{author}{\bibinfo{person}{J. Ménétrey}, \bibinfo{person}{C. Göttel}, \bibinfo{person}{M. Pasin}, \bibinfo{person}{P. Felber}, {and} \bibinfo{person}{V. Schiavoni}.} \bibinfo{year}{2022}\natexlab{}.
\newblock \bibinfo{title}{An Exploratory Study of Attestation Mechanisms for Trusted Execution Environments}.
\newblock
\showeprint[arxiv]{2204.06790}~[cs.CR]


\bibitem[Network(2025)]%
        {automata_dcap_attestation}
\bibfield{author}{\bibinfo{person}{Automata Network}.} \bibinfo{year}{2025}\natexlab{}.
\newblock \bibinfo{title}{Automata DCAP Attestation}.
\newblock
\urldef\tempurl%
\url{https://github.com/automata-network/automata-dcap-attestation}
\showURL{%
\tempurl}
\newblock
\shownote{Accessed: 2025-04-03}.


\bibitem[Perezvargas(2023)]%
        {microsoftConfidentialAzure}
\bibfield{author}{\bibinfo{person}{Caroline Perezvargas}.} \bibinfo{year}{2023}\natexlab{}.
\newblock \bibinfo{title}{Confidential VMs on Azure}.
\newblock
\newblock
\shownote{\url{https://techcommunity.microsoft.com/blog/windowsosplatform/confidential-vms-on-azure/3836282}}.


\bibitem[Rabimba et~al\mbox{.}(2021)]%
        {Rabimba_2021}
\bibfield{author}{\bibinfo{person}{Karanjai Rabimba}, \bibinfo{person}{Lei Xu}, \bibinfo{person}{Lin Chen}, \bibinfo{person}{Fengwei Zhang}, \bibinfo{person}{Zhimin Gao}, {and} \bibinfo{person}{Weidong Shi}.} \bibinfo{year}{2021}\natexlab{}.
\newblock \showarticletitle{Lessons Learned from Blockchain Applications of Trusted Execution Environments and Implications for Future Research}. In \bibinfo{booktitle}{\emph{Workshop on Hardware and Architectural Support for Security and Privacy}}. \bibinfo{publisher}{ACM}, \bibinfo{pages}{1–8}.
\newblock
\href{https://doi.org/10.1145/3505253.3505259}{doi:\nolinkurl{10.1145/3505253.3505259}}


\bibitem[Sardar et~al\mbox{.}(2021)]%
        {9448036}
\bibfield{author}{\bibinfo{person}{M.~U. Sardar}, \bibinfo{person}{S. Musaev}, {and} \bibinfo{person}{C. Fetzer}.} \bibinfo{year}{2021}\natexlab{}.
\newblock \showarticletitle{Demystifying Attestation in Intel Trust Domain Extensions via Formal Verification}.
\newblock \bibinfo{journal}{\emph{IEEE Access}}  \bibinfo{volume}{9} (\bibinfo{year}{2021}), \bibinfo{pages}{83067--83079}.
\newblock
\href{https://doi.org/10.1109/ACCESS.2021.3087421}{doi:\nolinkurl{10.1109/ACCESS.2021.3087421}}


\bibitem[SUSE(2025)]%
        {coconut-svsm2025Feb}
\bibfield{author}{\bibinfo{person}{SUSE}.} \bibinfo{year}{2025}\natexlab{}.
\newblock \bibinfo{title}{{svsm}}.
\newblock
\urldef\tempurl%
\url{https://github.com/coconut-svsm/svsm}
\showURL{%
\tempurl}
\newblock
\shownote{[Online; accessed 14. Feb. 2025]}.


\end{thebibliography}
\end{document}